\documentclass[aps,prb,reprint,superscriptaddress,10pt,longbibliography]{revtex4-2}
\usepackage[colorlinks,bookmarks=false,citecolor=blue,linkcolor=blue,urlcolor=blue]{hyperref}
\usepackage{amsmath, amssymb, graphicx, color}
\usepackage{verbatim}
\bibpunct{[}{]}{,}{n}{}{} 

\newcommand{\Det}{\mathop{\rm Det}\nolimits}
\newcommand{\Ree}{\mathop{\rm Re}\nolimits}
\newcommand{\Imm}{\mathop{\rm Im}\nolimits}
\newcommand{\sgn}{\mathop{\rm sgn}\nolimits}

\begin{document}

\title{Effective low-energy models for superconducting impurity systems}

\author{Vladislav Pokorn\'y}
\email{pokornyv@fzu.cz}
\affiliation{Institute of Physics, Czech Academy of Sciences, 
Na Slovance 2, CZ-18221 Praha 8, Czech Republic}

\author{Martin \v{Z}onda}
\email{martin.zonda@karlov.mff.cuni.cz}
\affiliation{Department of Condensed Matter Physics, Faculty of Mathematics and Physics, 
Charles University, Ke Karlovu 5, CZ-12116  Praha 2, Czech Republic}

\date{\today}

\begin{abstract}
We present two complementary methods to calculate the Andreev bound state energies of a single-level quantum dot connected to superconducting leads described by the superconducting impurity Anderson model. The first method, which is based on a mapping to a low-energy model, can be utilized to extract the Andreev bound state energies from finite-temperature, imaginary-time quantum Monte Carlo data without the necessity of any analytic continuation technique. The second method maps the full model on an exactly solvable superconducting atomic limit with renormalized parameters. As such, it represents a fast and reliable method for a quick scan of the parameter space. We demonstrate that after adding a simple band correction this method can provide predictions for measurable quantities, including the Josephson current, that are in a solid quantitative agreement with precise results obtained by the numerical renormalization group and quantum Monte Carlo.
\end{abstract}

\maketitle

\section{Introduction \label{Sec:Intro}}

Nanoscopic systems consisting of quantum dots coupled to superconducting leads have
attracted a lot of attention over the last few decades due to their possible applications
in quantum computing and sensor technologies
(for reviews, see~\cite{Wernsdorfer-2010,Rodero-2011,Meden-2019,Benito-2020}).
Several types of their experimental realizations are available.
One of them are single atoms or molecules deposited on the surface of a superconductor
and probed by (metallic or superconducting) scanning tunneling microscope
tip~\cite{Heinrich-2018,Kezilebieke-2018,Mier-2021}.
Another typical realizations involve short semiconducting nanowires, e.g., InAs or InSb,
connected to bulk superconducting
leads~\cite{vanDam-2006,Su-2017,Grove-Rasmussen-2018}.
As a result of the recent advances in fabrication
techniques, these devices allow for large control over the system parameters, e.g., via
different geometries of additional gates or by tuning the voltage gate and the
superconducting phase difference. As such, they present a rich playground allowing us to
investigate a multitude of physical phenomena including the supercurrent-carrying
Andreev bound states (ABS), which appear inside the superconducting gap induced on
the quantum dot.

Understanding the behavior of these states is crucial for the description of such
hybrid systems as they govern much of the transport properties. In addition, the
crossings of ABS at the Fermi energy mark quantum phase transitions (QPT), e.g., the
$0-\pi$ (singlet-doublet) transition known from single quantum dot systems~\cite{Oguri-2004}.
Considering the proposed applications, and with respect to future engineering,
it is, therefore, also crucial to develop practical and reliable methods for a correct
quantitative prediction of ABS.

Such superconducting hybrid systems are often reliably described by the
superconducting impurity Anderson model (SCIAM)~\cite{Luitz-2012}, which represents
single or multiple correlated quantum levels coupled to one or several
superconducting baths. For this model, a large variety of solvers emerged
over the years. They spread from  mappings to exactly solvable effective models, like
the superconducting atomic limit~\cite{Rozhkov-2000} or the zero-bandwidth
model~\cite{Grove-Rasmussen-2018}, through various diagrammatic perturbation
techniques including the Hartree-Fock approximation,
second-order perturbation theory~\cite{Vecino-2003,Zonda-2015,Zonda-2016},
non-crossing approximation~\cite{Ishizaka-1995,Clerk-2000}
and various advanced diagram resummation techniques~\cite{Janis-2021},
to heavy numerical methods, especially the numerical renormalization group (NRG)
method~\cite{Bulla-2008,Yoshioka-2000,Bauer-2007,Zitko-2015,Zitko-2016} and the
quantum Monte Carlo (QMC) in its various flavors including the
Hirsch-Fye method~\cite{Siano-2004}, and the continuous-time
interaction-expansion (CT-INT)~\cite{Luitz-2010,Luitz-2012,Delagrange-2015}
and hybridization-expansion (CT-HYB) \cite{Pokorny-2018,Pokorny-2021}
techniques.

Each of these methods has its advantages but also limitations that restrict
their applicability to certain regimes. For example, the diagrammatic
expansion techniques in the Coulomb interaction strength $U$ are fast,
simple and provide a reasonable solution in the weak and intermediate
interaction regime, but they are bound to situations where the ground
state is a singlet~\cite{Zonda-2015,Zonda-2016,Pokorny-2020}. This is
because the $U=0$ limit is always a singlet. Consequently, the doublet state
can not be reached by an adiabatic switch on of the interaction as
these two states are separated by a QPT.

A clear advantage of NRG, often the method of choice for SCIAM, is that it can
provide an unbiased solution at zero and low temperatures. However,
its numerical complexity grows exponentially with the number of channels
(i.e., terminals). Despite the recent advances~\cite{Zalom-2021,Zalom-2021-reentrant},
this still limits its applicability in case of complex setups.

On the other hand, the QMC methods are able to provide numerically exact solution
of SCIAM even for complicated devices, but they are bound to finite temperatures as they
are often formulated in the imaginary-time domain. Therefore, obtaining the spectral
function and the ABS energies requires performing an analytic continuation of
stochastic imaginary-time data to the real frequency domain, which is an
ill-defined problem~\cite{Jarrell-1996}.
While real-time implementations of QMC algorithms like the inchworm
method~\cite{Cohen-2015,Chen-2017} emerged recently, they have not yet been utilized
to solve superconducting models.

In this paper we introduce a method that allows us to extract the ABS energies directly from imaginary-time (or imaginary-frequency) QMC data by mapping the SCIAM on a low-energy model. That way the ill-defined analytic continuation can be avoided. The method is built on an older idea, which provides the microscopic basis for the Fermi liquid theory~\cite{Nozieres-1998}. In addition, we present a simple and reliable method based on the superconducting atomic limit called the generalized atomic limit (GAL). This method was originally utilized just to obtain the phase boundary between the $0$ and $\pi$ phases in a single quantum dot system. However, it is also able to provide the ABS energies with reasonable accuracy in a large part of the model parameter space, while being orders of magnitude less computationally expensive than NRG or QMC. Here we focus on the simple case of a single quantum dot connected to two superconducting leads and show how GAL can be improved even further by introducing a simple band correction. GAL for more complicated setups is presented elsewhere~\cite{Zonda-2023}.

The paper is organized as follows. In Sec.~\ref{Sec:Method} we present the SCIAM and the basic methods, which we later employ. We summarize the most important features of the superconducting atomic limit as they will prove to be useful in the next parts. Then we introduce the mapping on the low-energy model, which allows us to extract the ABS energies from imaginary-time QMC calculation without the need of analytic continuation. We also present a recipe on how to obtain GAL using a similar mapping. Although this recipe is far from rigorous derivation and it is guided mostly by comparison with numerically exact techniques, it provides a fast and simple method to study the behavior of SCIAM. In Sec.~\ref{Sec:Results} we present results of the two methods compared to NRG data for a simple case of a single quantum dot connected to two superconducting leads. We study the reliability of these methods by investigating the dependence of various properties on the interaction strength, temperature, phase difference, and the local energy level. We also discuss the fate of the second pair of ABS, which may be present in the $\pi$ phase of the model. Finally, in Sec.~\ref{Sec:Conclusions} we summarize the results and provide an outlook on the applicability for more complex setups. Furthermore, we discuss in Appendices~\ref{App:GALcalc} and~\ref{App:BandCorr} some technical details of the methods, which are omitted in the main text for the sake of readability.

\section{Model and method \label{Sec:Method}}

The SCIAM Hamiltonian of a single quantum dot connected to two superconducting BCS leads
reads
\begin{equation}
\label{Eq:HAMsciam}
\mathcal{H}=\mathcal{H}_{d}+\mathcal{H}_{U}+\sum_\alpha(\mathcal{H}^\alpha_{c}
+\mathcal{H}^\alpha_{\mathrm{hyb}}),\quad \alpha=L,R.
\end{equation}
The quantum dot is described as a single spinful atomic level,
\begin{equation}
\mathcal{H}_{d}=\varepsilon\sum_{\sigma}
d_{\sigma}^\dag d_{\sigma}^{\phantom{\dag}},
\end{equation}
with a local Coulomb interaction term that reads
\begin{equation}
\mathcal{H}_{U}=
U\left(d_{\uparrow}^\dag d_{\uparrow}^{\phantom{\dag}}-\frac{1}{2}\right)
\left(d_{\downarrow}^\dag d_{\downarrow}^{\phantom{\dag}}-\frac{1}{2}\right).
\end{equation}
Here $d_{\sigma}^\dag$ creates an electron with spin $\sigma$ and energy
$\varepsilon=\varepsilon_d+U/2$ on the quantum dot,
$\varepsilon_d$ is the local energy level
and $U$ is the repulsive on-site Coulomb
interaction. Hamiltonian of the superconducting lead $\alpha$ reads
\begin{equation}
\begin{aligned}
\mathcal{H}^\alpha_{c}&=
\sum_{\mathbf{k}\sigma}\varepsilon_{\mathbf{k}}
c_{\alpha\mathbf{k}\sigma}^\dag c_{\alpha\mathbf{k}\sigma}^{\phantom{\dag}} \\
&-\Delta\sum_\mathbf{k}(e^{i\varphi_\alpha}
c_{\alpha\mathbf{k}\uparrow}^\dag c_{\alpha\mathbf{-k}\downarrow}^\dag+\textrm{H.c.}),
\end{aligned}
\end{equation}
where $c_{\alpha\mathbf{k}\sigma}^\dag$ creates an electron with spin
$\sigma$ and energy $\varepsilon_{\mathbf{k}}$ in lead $\alpha$,
$\Delta e^{i\varphi_\alpha}=
g\langle c_{\alpha\mathbf{-k}\downarrow}^{\phantom{\dag}}
c_{\alpha\mathbf{k}\uparrow}^{\phantom{\dag}}\rangle$ is the BCS
superconducting order parameter with amplitude $\Delta$ and phase $\varphi$ and
$g$ is the attractive interaction strength in the leads.
We assume that the dispersion relation $\varepsilon_{\mathbf{k}}$ and the
amplitude $\Delta$ is the same for both leads (i.e., they are made from the same
material), but the superconducting phases $\varphi_\alpha$ can differ.
Finally, the coupling between the dot and the lead $\alpha$ is described by
\begin{equation}
\mathcal{H}^\alpha_{\mathrm{hyb}}=
-\sum_{\mathbf{k}\sigma}(V_{\alpha\mathbf{k}}
c_{\alpha\mathbf{k}\sigma}^\dag d_{\alpha\sigma}^{\phantom{\dag}}+\textrm{H.c.}),
\end{equation}
where $V_{\alpha\mathbf{k}}$ is the tunneling matrix element.

We define Nambu spinors for impurity and lead electrons,
$D^\dag=\left(d_{\uparrow}^\dag, d_{\downarrow}^{\phantom{\dag}}\right)$,
$C^\dag_{\alpha\mathbf{k}}=
\left(c_{\alpha\mathbf{k}\uparrow}^\dag,c_{\alpha\mathbf{-k}\downarrow}^{\phantom{\dag}}\right)$
and matrices
\begin{equation}
\begin{aligned}
E_{\alpha\mathbf{k}}&=
\begin{pmatrix}
\varepsilon_{\mathbf{k}} & -\Delta e^{i\varphi_\alpha} \\[0.3em]
-\Delta e^{-i\varphi_\alpha} & -\varepsilon_{\mathbf{-k}}
\end{pmatrix},\quad
E=
\begin{pmatrix}
\varepsilon & 0 \\[0.3em]
0 & -\varepsilon
\end{pmatrix}, \\
V_{\alpha\mathbf{k}}&=
\begin{pmatrix}
V_{\alpha\mathbf{k}} & 0 \\[0.3em]
0 & -V_{\alpha\mathbf{-k}}
\end{pmatrix}.
\end{aligned}
\end{equation}
The SCIAM Hamiltonian can be then rewritten, up to a constant term, as
\begin{equation}
\begin{aligned}
\mathcal{H}&=D^\dag E D+\mathcal{H}_{U}+\sum_{\alpha\mathbf{k}}
C_{\alpha\mathbf{k}}^\dag E_{\alpha\mathbf{k}}C_{\alpha\mathbf{k}}\\
&-\sum_{\alpha\mathbf{k}}(C_{\alpha\mathbf{k}}^\dag V_{\alpha\mathbf{k}}D
+\mathrm{H.c.}).
\end{aligned}
\end{equation}

Our main object of interest is the
impurity Green function $G(\tau)=-\langle\mathcal{T}_\tau[D(\tau)D^\dag(0)]\rangle$,
where $\mathcal{T}_\tau$ is the imaginary-time ordering operator.
As the superconducting correlations are already treated on the BCS level,
the lead degrees of freedom can be integrated out.
To avoid the complicated analytic structure of the Green function for a gapped
system, we resort to Matsubara (imaginary) frequency formalism for now.
We denote the non-interacting ($U=0$) Green function as $G_0$.
It reads
\begin{equation}
\begin{aligned}
G_0(i\omega_n)&=\int_{0}^{\beta}d\tau e^{i\omega_n\tau}G_0(\tau) \\
&=\Big[i\omega_n I_2-E-\sum_{\alpha}\Gamma_\alpha(i\omega_n)\Big]^{-1},
\end{aligned}
\end{equation}
where $\omega_n=(2n+1)\pi k_BT$ is the $n$th fermionic Matsubara frequency
at temperature $T$, $I_2$ is the $2\times 2$ unit matrix and
$\Gamma_\alpha(i\omega_n)$ is the hybridization function
between the lead $\alpha$ and the dot. It describes the hopping from
the impurity to the lead, the propagation through the
lead, and the hopping back to the impurity and can be written as
\begin{equation}
\Gamma_\alpha(i\omega_n)=\sum_{\mathbf{k}}
V^*_{\alpha\mathbf{k}}\mathcal{G}_{\alpha\mathbf{k}}(i\omega_n)V_{\alpha\mathbf{k}},
\end{equation}
where $\mathcal{G}_{\alpha\mathbf{k}}(i\omega_n)=[i\omega_n I_2-E_{\alpha\mathbf{k}}]^{-1}$
is the Green function of lead $\alpha$.
If we assume constant density of states in the band of half-width $W$,
$\rho(\varepsilon)=\Theta(\varepsilon^2-W^2)/2W$, we can transform the momentum
summation into an integral over energies. The hybridization function reads
\begin{equation}
\begin{aligned}
\Gamma_\alpha(i\omega_n)
&=-\frac{\Gamma_\alpha w(i\omega_n)}{\sqrt{\omega_n^2+\Delta^2}}
\begin{pmatrix}
i\omega_n & \Delta e^{i\varphi_\alpha} \\[0.3em]
\Delta e^{-i\varphi_\alpha} & i\omega_n
\end{pmatrix},
\end{aligned}
\end{equation}
where we defined the tunneling rates $\Gamma_\alpha=\pi|V_{\alpha}|^2/(2W)$
and
\begin{equation}
w(i\omega_n)=\frac{2}{\pi}\arctan\left(\frac{W}{\sqrt{\omega_n^2+\Delta^2}}\right)
\end{equation}
is the correction to finite bandwidth that approaches unity for $W\rightarrow\infty$.
The non-interacting impurity Green function then reads
\begin{equation}
\label{Eq:nonintGF}
\begin{aligned}
&G^{-1}_0(i\omega_n)=\\
&
\begin{pmatrix}
i\omega_n[1+s(i\omega_n)]-\varepsilon & \Delta_\varphi(i\omega_n) \\[0.3em]
\Delta^*_\varphi(i\omega_n) & i\omega_n[1+s(i\omega_n)]+\varepsilon
\end{pmatrix}.
\end{aligned}
\end{equation}
Here we denoted
\begin{equation}
\label{Eq:HybSE}
s(i\omega_n)=\frac{\Gamma w(i\omega_n)}{\sqrt{\Delta^2+\omega_n^2}},\quad
\Delta_\varphi(i\omega_n)=\frac{\Delta\Gamma_\varphi w(i\omega_n)}
{\sqrt{\Delta^2+\omega_n^2}},
\end{equation}
$\Gamma=\Gamma_L+\Gamma_R$ and
$\Gamma_\varphi=\Gamma_Le^{i\varphi_L}+\Gamma_Re^{i\varphi_R}$.
We emphasize that, due to the gauge invariance,
all physical observables can depend only on the phase difference
$\varphi=\varphi_L-\varphi_R$ and not on the values of the individual
phases~\cite{Meden-2019}.
This property can be utilized to keep the off-diagonal term $\Delta_\varphi$
real by a proper shift of both superconducting phases,
$\varphi_\alpha\rightarrow\varphi_\alpha+\varphi_s$.
We also note that any setup with asymmetric coupling $\Gamma_L\neq\Gamma_R$
can be easily transformed to the symmetric case~\cite{Kadlecova-2017},
for which $\Gamma_\varphi=\Gamma\cos(\varphi/2)$.

The non-interacting Green function can be straightforwardly continued to real
frequencies, $i\omega_n\rightarrow\omega\pm i0$. The frequency-dependent
factors in Eq.~\eqref{Eq:HybSE} then read~\cite{Zonda-2015}
\begin{equation}
\begin{aligned}
s(\omega\pm i0)&=\Gamma x(\omega\pm i0), \\
\Delta_\varphi(\omega\pm i0)&=\Delta\Gamma_\varphi x(\omega\pm i0),
\end{aligned}
\end{equation}
where
\begin{equation}
\label{Eq:xcont}
\begin{aligned}
x(\omega\pm i0)&=\pm\frac{i\sgn{\omega}}{\sqrt{\omega^2-\Delta^2}},\quad |\omega|>\Delta,\\
x(\omega\pm i0)&=\frac{1}{\sqrt{\Delta^2-\omega^2}},\quad |\omega|<\Delta.
\end{aligned}
\end{equation}
Finally, the correction to finite bandwidth reads
\begin{equation}
w(\omega\pm i0)=\frac{2}{\pi}\arctan[Wx(\omega\pm i0)].
\end{equation}
For the sake of simplicity we drop this factor from the equations.
It can be reintroduced later, if needed, by scaling $\Gamma\rightarrow\Gamma w(\omega)$
and $\Gamma_\varphi\rightarrow\Gamma_\varphi w(\omega)$ in the final expressions.

The symmetry relations for the diagonal (normal)
and off-diagonal (anomalous) elements of the Green function
in the real frequency domain read
\begin{equation}
\begin{aligned}
G_{22}(\omega+i0)&=-G_{11}(-\omega-i0)=-G^*_{11}(-\omega+i0), \\
G_{21}(\omega+i0)&=G_{12}(-\omega-i0)=G^*_{12}(-\omega+i0).
\end{aligned}
\end{equation}
They reduce the number of independent elements to two which we mark
$G_n\equiv G_{11}$ and $G_a\equiv G_{12}$. Moreover, for real
$\Delta_\varphi$ the anomalous elements are even functions
of the frequency and therefore
$G_{12}(\omega+i0)=G_{21}(\omega+i0)\equiv G_a(\omega+i0)$.

The knowledge of the anomalous part of the impurity Green function also allows us to
calculate the equilibrium, dc Josephson current driven by the phase
difference $\varphi$. It can be derived from the
Heisenberg equation of motion and reads~\cite{Meden-2019}
\begin{equation}
\label{Eq:JC}
J_\alpha=\frac{J_0}{\beta}\sum_n\frac{\Gamma_\alpha}{\sqrt{\Delta^2+\omega_n^2}}
\Imm\left[G_a(i\omega_n)e^{-i\varphi_\alpha}\right],
\end{equation}
where $\alpha=L,R$ marks the direction of the current and $J_0=e\Delta/\hbar$.
The analytic continuation of this formula to the real frequency axis can be found, e.g.,
in Ref.~\cite{Zalom-2021}.

\subsection{Superconducting atomic limit \label{SSec:AtomicLimit}}
The basic properties of SCIAM can be illustrated on the
analytically solvable case of $\Delta\rightarrow\infty$.
This regime is usually called the superconducting atomic limit and
in order to obtain a non-trivial atomic model, the limit of
$W\rightarrow\infty$ must be taken first otherwise the proximity effect
would be lost. The non-interacting Green function then reads
\begin{equation}
 \label{Eq:G0atomic}
 G^{-1}_{\infty0}(\omega)=
 \begin{pmatrix}
  \omega-\varepsilon & \Gamma_\varphi \\[0.3em]
  \Gamma_\varphi & \omega+\varepsilon
 \end{pmatrix}
\end{equation}
and SCIAM reduces to a local atomic model with off-diagonal on-site
term~\cite{Rozhkov-2000}.
The Hamiltonian in this case reads
\begin{equation}
 \label{Eq:HamAL}
 \mathcal{H}_{\infty}=\mathcal{H}_{d}+\mathcal{H}_{U}
 -(\Gamma_\varphi d_{\uparrow}^\dag d_{\downarrow}^\dag+\mathrm{H.c.}).
\end{equation}
This limit was already abundantly discussed in
literature~\cite{Vecino-2003,Bauer-2007,Tanaka-2007,Meng-2009,Janis-2021} so we just briefly
summarize the results important for this paper.

The behavior of ABS and the basic physics behind the $0-\pi$ QPT can
be demonstrated on the energy spectrum of the atomic model.
The eigenspectrum of this model consists of a Kramers doublet with energy
$\varepsilon_d$ and a pair of singlets with energies $E_\pm=\varepsilon\pm R$,
where we introduced $R=\sqrt{\Gamma_\varphi^2+\varepsilon^2}$.
The number of states in the excitation spectrum then depends on the parity of
the ground state. For singlet ground state the
excitation spectrum consists of two ABS that correspond to transitions
between the lower singlet $E_-$ and the doublet, $E_0=\pm(-U/2+R)$.
The singlet-singlet transition violates the $\Delta s_z=\pm 1/2$ selection
rule and does not contribute to the single-particle spectrum.
For the doublet ground state we obtain two pairs of energies
$E_{0+}=\pm(U/2+R)$ and $E_{0-}=\pm(U/2-R)$.

The normal and anomalous elements of the atomic Green function
$G_\infty(\omega)$ in the singlet
phase read
\begin{equation}
\begin{aligned}
G_{sn}(\omega)&=\frac{1}{2R}
\left(\frac{R-\varepsilon}{\omega+E_0}+\frac{R+\varepsilon}{\omega-E_0}\right), \\
G_{sa}(\omega)&=\frac{\Gamma_\varphi}{2R}\left(\frac{1}{\omega+E_0}-\frac{1}{\omega-E_0}\right).
\end{aligned}
\end{equation}
Note that at half filling
($\varepsilon=0$), all weights of the ABS equal $1/2$.
The electron density
$n=\sum_\sigma\langle d^\dag_\sigma d^{\phantom{\dag}}_\sigma\rangle$
and the induced pairing
$\nu=\langle d_\downarrow d_\uparrow\rangle$
are at zero temperature given by the weight of the state below the Fermi
energy, $n=1-\varepsilon/R$ and $\nu=\Gamma_\varphi/(2R)$.

In the doublet phase the elements of the Green function read
\begin{equation}
\begin{aligned}
&G_{dn}(\omega)=\\
&\frac{1}{4R}
\left(\frac{R+\varepsilon}{\omega+E_{0-}}+\frac{R-\varepsilon}{\omega+E_{0+}}
+\frac{R+\varepsilon}{\omega-E_{0+}}+\frac{R-\varepsilon}{\omega-E_{0-}}\right), \\
&G_{da}(\omega)=\\
&\frac{\Gamma_\varphi}{4R}\left(\frac{1}{\omega+E_{0-}}-\frac{1}{\omega+E_{0+}}
+\frac{1}{\omega-E_{0+}}-\frac{1}{\omega-E_{0-}}\right),
\end{aligned}
\end{equation}
from which we obtain that $n=1$ and $\nu=0$ at zero temperature
for all parameters as the two contributions to the induced pairing
cancel each other out.

The formula for the zero-temperature Josephson current~\eqref{Eq:JC}
reduces in the atomic limit to $J=(2e/\hbar)\partial E_g/\partial \varphi$,
where $E_g$ is the energy of the ground state~\cite{Meden-2019}.
If we assume $\Gamma_L=\Gamma_R=\Gamma/2$,
it reads $J=J_0\Gamma^2\sin \varphi/(2R)$ in the $0$ phase and $J=0$ in the
$\pi$ phase as its ground state energy $E_g=\varepsilon_d$ is
independent of the phase difference.

The boundary between the $0$ phase with singlet ground state and the $\pi$ phase
with doublet ground state is marked by the
crossing of ABS at the Fermi energy and therefore it is given by
the condition $E=0$, i.e., $R=U/2$. As for $R>U/2$ the system is in the $0$
phase, the non-interacting case is always a singlet
(except for $\varepsilon=0$ and $\varphi=\pi$, which is a transition point).

We can also formally define the self-energy in the superconducting atomic limit,
$\Sigma_\infty(\omega)=G^{-1}_{\infty 0}(\omega)-G_\infty^{-1}(\omega)$.
In the $0$ phase both the non-interacting and the interacting Green
function have two poles and the self-energy is a simple real shift
of the energies, which resembles the Hartree-Fock solution,
\begin{equation}
\label{Eq:SEatS}
\Sigma_{sn}=\frac{Un}{2},
\qquad\Sigma_{sa}=U\nu.
\end{equation}
For the $\pi$ phase the situation is more complicated as the
non-interacting and the interacting Green functions have different numbers
of poles. The self-energy then also has two poles at $\pm R$ and
reads
\begin{equation}
\label{Eq:SEatD}
 \begin{aligned}
 \Sigma_{dn}(\omega)&=\frac{U}{2}+\frac{U^2}{8R}
 \left[\frac{R-\varepsilon}{\omega+R}+\frac{R+\varepsilon}{\omega-R}\right]\\
 &=
 \frac{U}{2}+\frac{U^2}{4}G_{sn,0}(\omega), \\
 \Sigma_{da}(\omega)&=\frac{U^2\Gamma_\varphi}{8R}
 \left[\frac{1}{\omega+R}-\frac{1}{\omega-R}\right]=
 \frac{U^2}{4}G_{sa,0}(\omega),
 \end{aligned}
\end{equation}
where $G_{sn,0}$ and $G_{sa,0}$ are the normal and anomalous elements of $G_{\infty 0}$.
Note that the term $U/2$ in the normal part just compensates for the definition
of the energy level $\varepsilon=\varepsilon_d+U/2$ in the non-interacting
Green function and the non-trivial part of the self-energy is of second order
in the interaction strength.

\subsection{Low-energy model \label{SSec:MethodLowEn}}
The superconducting atomic limit provides a qualitatively correct solution,
including the behavior around the $0-\pi$ QPT, but fails
to provide quantitatively reasonable results due to the missing band
contributions. This hints that most
of the physical properties are governed by the behavior of the ABS
while the incoherent band states above $\Delta$ cause
the renormalization of the energy.
Therefore, we separate the Green function into the low- and
high-energy parts. We can write the exact impurity Green function as
\begin{equation}
\label{Eq:FullGF}
\begin{aligned}
&G^{-1}(\omega)=G_0^{-1}(\omega)-\Sigma(\omega)=\\
&\begin{pmatrix}
\omega[1\!+\!s(\omega)]-\varepsilon-\Sigma_{n}(\omega) &
\Delta_\varphi(\omega)-\Sigma_{a}(\omega) \\[0.3em]
\Delta_\varphi(\omega)-\Sigma_{a}(\omega) &
\omega[1\!+\!s(\omega)]+\varepsilon+\Sigma^*_{n}(-\omega)
\end{pmatrix},
\end{aligned}
\end{equation}
where $\Sigma(\omega)$ is the exact self-energy in Nambu formalism,
\begin{equation}
\Sigma(\omega)=
\begin{pmatrix}
\Sigma_n(\omega) & \Sigma_a(\omega) \\
\Sigma_a(\omega) & -\Sigma^*_n(-\omega)
\end{pmatrix}.
\end{equation}
The expansion around $\omega=0$ of the frequency-dependent terms reads
\begin{equation}
\label{Eq:LowFreqExp}
\begin{aligned}
s(\omega)&=\frac{\Gamma}{\Delta}+
\frac{\Gamma}{2\Delta^3}\omega^2+\mathcal{O}(\omega^4), \\
\Delta_\varphi(\omega)&=\Gamma_\varphi+
\frac{\Gamma_\varphi}{2\Delta^2}\omega^2+\mathcal{O}(\omega^4), \\
\Sigma_j(\omega)&=\Sigma_j(0)+\omega\frac{\partial\Sigma_j}{\partial\omega}\Bigr|_0
+\omega^2\frac{1}{2}\frac{\partial^2\Sigma_j}{\partial\omega^2}\Bigr|_0
+\mathcal{O}(\omega^3),
\end{aligned}
\end{equation}
$j=n,a$.
The first derivative of the anomalous part $\partial\Sigma_a/\partial\omega|_0$
is always zero due to symmetry reasons.
The Green function can be thus written in a form
\begin{equation}
\label{Eq:GreenLowEn}
\begin{aligned}
G^{-1}(\omega)&=Z^{-1}
\begin{pmatrix}
\omega-\tilde{\varepsilon}-\tilde{\Sigma}_{n}(0) &
\tilde{\Gamma}_\varphi-\tilde{\Sigma}_{a}(0) \\[0.3em]
\tilde{\Gamma}_\varphi-\tilde{\Sigma}_{a}(0) &
\omega+\tilde{\varepsilon}+\tilde{\Sigma}^*_{n}(0)
\end{pmatrix} \\
&+\mathcal{C}(\omega)=
Z^{-1}[\tilde{G}^{-1}(\omega)+\tilde{\mathcal{C}}(\omega)],
\end{aligned}
\end{equation}
where
\begin{equation}
\label{Eq:Zfac}
Z^{-1}=1+\frac{\Gamma}{\Delta}-\frac{\partial \Sigma_{n}}{\partial \omega}\Bigr|_0
\end{equation}
is the renormalization factor,
$\tilde{\varepsilon}=Z\varepsilon$,
$\tilde{\Gamma}_\varphi=Z\Gamma_\varphi$,
$\tilde{\Sigma}=Z\Sigma$,
$\tilde{\mathcal{C}}=Z\mathcal{C}$ and
$\mathcal{C}$ is the correction, which contains all the higher-order contributions
to $s(\omega)$, $\Delta_\varphi(\omega)$ and $\Sigma(\omega)$,
including the incoherent band states.

The low-energy part resembles the non-interacting
Green function in the atomic limit, Eq.~\eqref{Eq:G0atomic},
with renormalized parameters. As the ground state in the
non-interacting case is always a singlet, this model can describe only
one pair of ABS even in the $\pi$ phase. If we neglect the
correction $\mathcal{C}$, we get simple formulas for the ABS energies as
zeros of the determinant $\Det[\tilde{G}^{-1}(\omega)]$,
which read
\begin{equation}
\label{Eq:ABSbareCTHYB}
\begin{aligned}
E_0&=\pm\sqrt{[\tilde{\varepsilon}+\tilde{\Sigma}_n(0)]^2+
[\tilde{\Gamma}_\varphi-\tilde{\Sigma}_a(0)]^2} \\
&=\pm Z\sqrt{[\varepsilon+\Sigma_n(0)]^2+[\Gamma_\varphi-\Sigma_a(0)]^2}.
\end{aligned}
\end{equation}
Nevertheless, including the correction $\mathcal{C}$ leads to a better
approximation. In the exact limit it should contain all the higher-order
contributions to the self-energy $\Sigma(\omega)$, which are generally
not known. If we neglect these contributions, the non-interacting part
of the correction reads
\begin{equation}
\label{Eq:Corr}
\mathcal{C}(\omega)=p(\omega)
\begin{pmatrix}
\Gamma\omega/\Delta & \Gamma_\varphi \\[0.3em]
\Gamma_\varphi & \Gamma\omega/\Delta
\end{pmatrix},
\end{equation}
where $p(\omega)=\Delta/x(\omega)-1$
with $x(\omega)$ given by Eq.~\eqref{Eq:xcont}.
This correction is most important in the weakly interacting regime $U<\Gamma$,
where the behavior of the system is governed
mostly by the hybridization function and for large values of the ABS energy
approaching the gap edge $\Delta$ where the low-energy model naturally fails.

As such correction vanishes at $\omega=0$, it has no effect
on the position of the QPT, which can be obtained from Eq.~\eqref{Eq:ABSbareCTHYB}
as the zero of the right-hand side. This means it depends solely on the model
parameters and the value of the self-energy at zero frequency.
The equation $E_0=0$ has two solutions and reads
\begin{equation}
\Gamma_\varphi-\Sigma_a(0)=\pm[\varepsilon+\Sigma_n(0)].
\end{equation}
The existence of two solutions reflects the electron-hole symmetry,
which implies that if there is a QPT at $\varepsilon=\varepsilon_c$ there
is also a QPT at $\varepsilon=-\varepsilon_c$.

It is possible to further, systematically improve the result of this method
by considering more terms of the frequency expansion~\eqref{Eq:LowFreqExp},
which is useful in the case of strong Coulomb interaction.
Including the second term in the expansion we obtain
\begin{equation}
\label{Eq:ExtLowEnergy}
\begin{aligned}
ZG_n^{-1}&=
-\frac{\tilde{\Sigma}^{\prime\prime}_{n}(0)}{2}\omega^2
+\omega-\tilde{\varepsilon}-\tilde{\Sigma}_{n}(0), \\
ZG_a^{-1}&=
\frac{1}{2}\left(\frac{\tilde{\Gamma}_\varphi}{\Delta^2}
-\tilde{\Sigma}^{\prime\prime}_{a}(0)\right)\omega^2
+\tilde{\Gamma}_\varphi-\tilde{\Sigma}_{a}(0),
\end{aligned}
\end{equation}
where we marked the first and second frequency derivatives of the self-energy
at zero as $\Sigma^{\prime}_j(0)$ and $\Sigma^{\prime\prime}_j(0)$.
There are four zeros of the determinant in this case, which represent four
bound states. This approach, however, does not solve the
above-mentioned problem with the missing ABS in the $\pi$ phase
as the two additional solutions lie always above the gap edge $\Delta$
as we discuss later.

This low-energy model is useful to overcome the notorious disadvantage
of the imaginary-time QMC methods in which the spectral function
can be obtained only via analytic continuation of the imaginary-time
or imaginary-frequency stochastic data, which is a known ill-defined problem
due to the exponential nature of the transformation kernel~\cite{Jarrell-1996}.
On the other hand, the values $\Sigma_j(0)$
and the first few derivatives can be obtained
from imaginary-frequency data using the Cauchy-Riemann equations.
For $T\rightarrow 0$ and $z=\omega+i\omega_n$ we obtain
\begin{equation}
\label{Eq:CauchyRiemann1}
\frac{\partial \Ree\Sigma_j(z)}{\partial \omega}=
\frac{\partial \Imm\Sigma_j(z)}{\partial \omega_n},\quad
\frac{\partial \Ree\Sigma_j(z)}{\partial \omega_n}=
-\frac{\partial \Imm\Sigma_j(z)}{\partial \omega}.
\end{equation}
A similar approach was already utilized to obtain the Fermi liquid parameters from QMC simulations of metallic systems~\cite{Arsenault-2012}. The formula for the second derivative of the real part reads
\begin{equation}
\label{Eq:CauchyRiemann2}
\frac{\partial^2 \Ree\Sigma_j(z)}{\partial \omega^2}=
-\frac{\partial^2 \Ree\Sigma_j(z)}{\partial \omega_n^2}.
\end{equation}
At finite temperatures the derivatives can be approximated by finite differences, $\Sigma'(0)\approx\Sigma(i\omega_0)/\omega_0$ where $\omega_0=\pi k_BT$ is the first positive Matsubara frequency. Similarly, the second derivative can be calculated from the first two positive frequencies.

\begin{figure}[htb]
\includegraphics[width=\columnwidth]{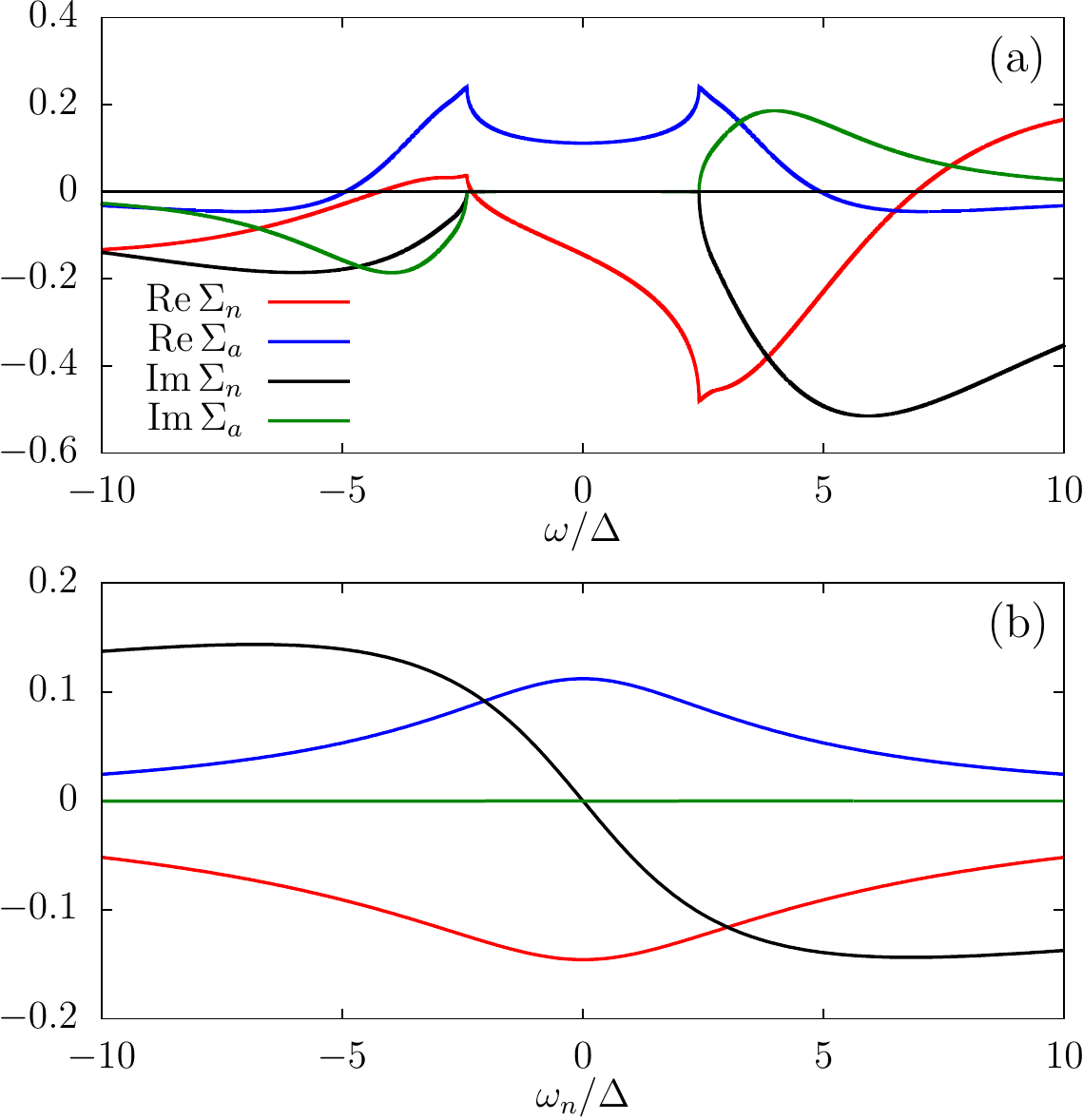}
\caption{Normal and anomalous components of the self-energy calculated using the second-order perturbation theory for $U=4\Delta$, $\Gamma=2\Delta$, $\varepsilon=2\Delta$, $\varphi=0$ and $T=0$ along the real (a) and imaginary frequency axis (b).
\label{Fig:SE_2pt}}
\end{figure}

To better illustrate relations~\eqref{Eq:CauchyRiemann1} and~\eqref{Eq:CauchyRiemann2}, we plotted in Fig.~\ref{Fig:SE_2pt}
the self-energy in both real and Matsubara frequency domain calculated at zero-temperature using the second-order perturbation theory~\cite{Zonda-2015} for $U=4\Delta$, $\Gamma=2\Delta$, $\varepsilon=2\Delta$ and $\varphi=0$.
Panel (a) shows the normal and the anomalous self-energy along the real frequency axis. Both imaginary parts contain a gap around the Fermi energy. As a result, the values and all derivatives at $\omega=0$ are real. The first derivative $\Sigma^\prime_n(0)$ is always non-positive (zero only for $U=0$) while $\Sigma^\prime_a(0)=0$ for symmetry reasons. The second derivative
$\Sigma^{\prime\prime}_n(0)$ is negative (positive)  for $\varepsilon>0$ ($\varepsilon<0$) and zero at half-filling, while $\Sigma^{\prime\prime}_a(0)\geq 0$ for all parameters (zero only for $\Delta=0$).
Panel (b) shows the same functions along the imaginary frequency axis. Here the derivative $\Sigma^\prime_n(0)$ is pure imaginary and matches the value of the derivative of the real part along the real axis according to Eq.~\eqref{Eq:CauchyRiemann1}, while both values of the second derivatives along the imaginary axis
are real and match the second derivatives along real axis but with opposite signs as given by~\eqref{Eq:CauchyRiemann2}.

\subsection{GAL \label{SSec:GAL}}
The GAL was introduced in Ref.~\cite{Zonda-2015}
as a simple formula for the position of the $0-\pi$ transition,
which gives a remarkably good agreement with the NRG in the
vicinity of half-filling. It was derived from the Hartree-Fock
result for the $0$ phase by neglecting the continuous band contribution
to the Green function, which is largely overestimated in the
Hartree-Fock treatment. This method was later modified for situations away
from half filling by fitting the NRG data~\cite{Kadlecova-2019}.

Here we present an approach motivated by the above-mentioned low-energy
construction, which results in the same formula for the $0-\pi$ transition as GAL
and also provides ABS energies as well as other model parameters.
The resulting formulas then represent a fast and reliable solver for SCIAM,
which can be used to scan the parameter space of the model before the computationally
more expensive methods like QMC or NRG are employed.

As the starting point we calculate the low-energy limit of the non-interacting Green
function~\eqref{Eq:nonintGF} to obtain the appropriate scaling of the model parameters,
\begin{equation}
\label{Eq:GFgalZero}
\begin{aligned}
G_0^{-1}(\omega)&\approx q^{-1}
\begin{pmatrix}
\omega-\tilde{\varepsilon}_\mu &
\tilde{\Gamma}_\varphi \\[0.3em]
\tilde{\Gamma}_\varphi &
\omega+\tilde{\varepsilon}_\mu
\end{pmatrix} + \mathcal{C}(\omega)\\
&=q^{-1}\tilde{G}_0^{-1}(\omega)+\mathcal{C}(\omega),
\end{aligned}
\end{equation}
where $q=(1+\Gamma/\Delta)^{-1}\leq 1$ is a renormalization factor to the finite
gap
$\Delta$~\footnote{
We note that a similar scaling procedure, but in different context,
was used by Kurilovich et al.~\cite{Kurilovich-2021} to calculate the
linear microwave response of a superconducting quantum dot in the $0$ phase.
},
$\mathcal{C}$ is the band correction given by Eq.~\eqref{Eq:Corr},
$\tilde{\varepsilon}_\mu=q\varepsilon_\mu$,
$\tilde{\Gamma}_\varphi=q\Gamma_\varphi$,
$\varepsilon_\mu=\varepsilon_d-\mu_s$ and $\mu_s$ is a
a yet arbitrary shift of the chemical potential, which guarantees that $\tilde{\varepsilon}_\mu=0$
corresponds to the electron-hole symmetric case as we discuss later.
The Green function $\tilde{G}_0$ has the same structure as in the superconducting
atomic limit and hence it corresponds to an auxiliary non-interacting problem
described by a Hamiltonian, which reads
\begin{equation}
\tilde{\mathcal{H}}_{\infty0}=
\sum_{\sigma}\tilde{\varepsilon}_\mu
\tilde{d}_{\sigma}^\dag \tilde{d}_{\sigma}^{\phantom{\dag}}
-(\tilde{\Gamma}_\varphi \tilde{d}_{\uparrow}^\dag \tilde{d}_{\downarrow}^\dag+\mathrm{H.c.}).
\end{equation}
Now we utilize our knowledge of the solution of the interacting problem in the
superconducting atomic limit, which is described by Hamiltonian
\begin{equation}
\label{Eq:HamGALFull}
\tilde{\mathcal{H}}_\infty=\tilde{\mathcal{H}}_{\infty0}
+\tilde{U}\left(\tilde{d}_{\uparrow}^\dag \tilde{d}_{\uparrow}^{\phantom{\dag}}-\frac{1}{2}\right)
\left(\tilde{d}_{\downarrow}^\dag \tilde{d}_{\downarrow}^{\phantom{\dag}}-\frac{1}{2}\right)
\end{equation}
and we replace the exact impurity Green function with the Green function in the
atomic limit with scaled parameters.
The approximation we made here is that we replaced the exact self-energy
$\Sigma(\omega)$ in the full impurity Green function~\eqref{Eq:FullGF}
by the scaled self-energy in the atomic limit $\Sigma_\infty(\omega)$
given by Eqs.~\eqref{Eq:SEatS} and~\eqref{Eq:SEatD},
\begin{equation}
\Sigma(\omega;\Delta,\varphi,\varepsilon,\Gamma,U)
\approx q^{-1}\Sigma_\infty(\omega;\Delta,\varphi,\tilde{\varepsilon},\tilde{\Gamma},\tilde{U}).
\end{equation}
Note that we did not yet specify the relation between $\tilde{U}$ and $U$.
The impurity Green function now reads
\begin{equation}
\label{Eq:GFgalFull}
\begin{aligned}
G^{-1}(\omega)&=G_0^{-1}(\omega)-\Sigma(\omega) \\
&\approx q^{-1}[\tilde{G}_0^{-1}(\omega)-\tilde{\Sigma}_\infty(\omega)
+\tilde{\mathcal{C}}(\omega)] \\
&=q^{-1}[\tilde{G}^{-1}(\omega)+\tilde{\mathcal{C}}(\omega)],
\end{aligned}
\end{equation}
where $\tilde{\Sigma}_\infty=q\Sigma_\infty$, $\tilde{\mathcal{C}}=q\mathcal{C}$
and $G_0$ is the non-interacting Green function given by Eq.~\eqref{Eq:nonintGF}.
Note that we ignored the frequency dependence of the self-energy while defining $q$.
This approach is therefore well justified only in the $0$ phase where the self-energy
is static.

Let us note that, similarly to the case of the Landau Fermi liquid,
the Green function~\eqref{Eq:GFgalFull} without the correction $\mathcal{C}$
does not describe a whole particle. The leading order in asymptotic expansion
of the diagonal element reads $G_n(\omega)\sim q/\omega$ and hence it
describes a quasiparticle with non-canonical anticommutation relation
$[d^{\phantom{\dag}}_\sigma, d^\dag_{\sigma^\prime}]_+=
q\delta_{\sigma\sigma^\prime}$. Therefore, the concept of a half-filled
band is misleading and $\mu_s\neq -U/2$ in the electron-hole symmetric case.
On the other hand, the Green function
$\tilde{G}(\omega)=[\tilde{G}_0^{-1}(\omega)-\tilde{\Sigma}_\infty(\omega)]^{-1}$
has the correct asymptotics as it corresponds to the atomic model~\eqref{Eq:HamGALFull}.

We still need to specify the values of $\tilde{U}$ and $\mu_s$.
They can be both obtained from the exact form of self-energy in the atomic
limit in the $0$ phase, Eq.~\eqref{Eq:SEatS}, as discussed in detail in
Appendix.~\ref{App:GALcalc}. We obtain
\begin{equation}
\label{Eq:GAL_U}
\tilde{U}=q^2U,\qquad \mu_s=-qU/2.
\end{equation}
The scaling of the energy levels follows
\begin{equation}
\tilde{\varepsilon}_\mu=q\varepsilon_\mu
=q\left(\varepsilon_d+\frac{qU}{2}\right)=\tilde{\varepsilon}_d+\frac{\tilde{U}}{2}
\end{equation}
and we drop the subscript $\mu$ from now on.

An alternative way to obtain the scaling of the interaction strength is to
formally redefine the creation and annihilation operators,
$\tilde{d}^{\phantom{\dag}}_\alpha=\sqrt{q}d^{\phantom{\dag}}_\alpha$
and $\tilde{d}^\dag_\alpha=\sqrt{q}d^\dag_\alpha$ so they obey standard anticommutation
relations. Inserting them into the Hamiltonian in the superconducting atomic limit,
Eq.~\eqref{Eq:HamAL}, we obtain the Hamiltonian of the auxiliary problem~\eqref{Eq:HamGALFull},
with the same scaling of the parameters as before,
$\tilde{\varepsilon}=q\varepsilon$,
$\tilde{\Gamma}_\varphi=q\Gamma_\varphi$
and $\tilde{U}=q^2U$.

Nevertheless, comparison of GAL results with NRG show good
agreement only in the vicinity of half filling. Detailed analysis of the data
show that a much better agreement can be obtained by introducing an
additional scaling of the local energy level, which follows~\cite{Kadlecova-2019}
\begin{equation}
\tilde{\varepsilon}\rightarrow q\sqrt{1+\frac{2\tilde{\Gamma}}{\tilde{U}}}\tilde{\varepsilon}.
\end{equation}
This scaling was obtained by fitting the NRG data for $\varphi=0$
and later proven to work for arbitrary value of the phase difference.
We denote this modified method as modified GAL (MGAL).
Unfortunately, the microscopic origin of this modification is still unknown and its
derivation would require a more rigorous treatment of the interaction part of the
Hamiltonian than the one we present in this paper.

\section{Results \label{Sec:Results}}

All CT-HYB calculations were performed using the TRIQS/CTHYB 3.0.1 solver~\cite{Seth-2016}. We set $W=100\Delta$ and the cutoff in Matsubara frequencies $\omega_{max}=200\Delta$. As the SCIAM Hamiltonian, Eq.~\eqref{Eq:HAMsciam}, is non-conserving, we perform a canonical electron-hole transformation in the spin-down segment of the Hilbert space to transform SCIAM into the standard impurity Anderson model with negative interaction strength $U$, as explained in detail in, e.g., Ref.~\cite{Pokorny-2018}. Calculations were performed using 288 CPU cores, $2\times 10^6-10^7$ QMC measurements per core.
We encountered no fermionic sign problem during the calculations. The total charge $n$ and the induced pairing $\nu$ were evaluated by measuring the impurity density matrix. The self-energy $\Sigma(i\omega_n)$ was obtained from the measured impurity Green function via the Dyson equation. Most of the data were calculated at temperature $k_BT=0.05\Delta$, which, e.g., for an aluminum electrode with $\Delta\approx150\mu$eV corresponds to $T\approx 77$mK. All calculations included the band correction $\mathcal{C}$.

GAL calculations were performed using a Python code based on the exact diagonalization solver for the atomic problem as implemented in the TRIQS libraries~\cite{Parcollet-2015}. The calculations were performed on a standard PC as a single data point can be calculated within a few seconds. Two versions of this method were employed, one that ignores the effects of the band correction $\mathcal{C}$ (GAL) and one that includes the correction (GAL+$\mathcal{C}$). The effects of $\mathcal{C}$ are discussed in detail in Appendix~\ref{App:BandCorr}.

Both zero-temperature and finite-temperature NRG data were used as a benchmark for our results. All NRG results were calculated via the NRG Ljubljana package~\cite{Zitko-Ljubljana} for $W=100\Delta$. In the case of single channel calculations ($\varphi = 0$) a logarithmic discretization parameter $\lambda=2$ or lower was used, the SPSU2 symmetry was utilized with minimal number of kept states set to 2000. In the case of two channel calculations ($\varphi \neq 0$) we used $\lambda=4$.

\subsection{Effect of interaction strength \label{SSec:ResU}}

\begin{figure*}[htb]
\includegraphics[width=2.0\columnwidth]{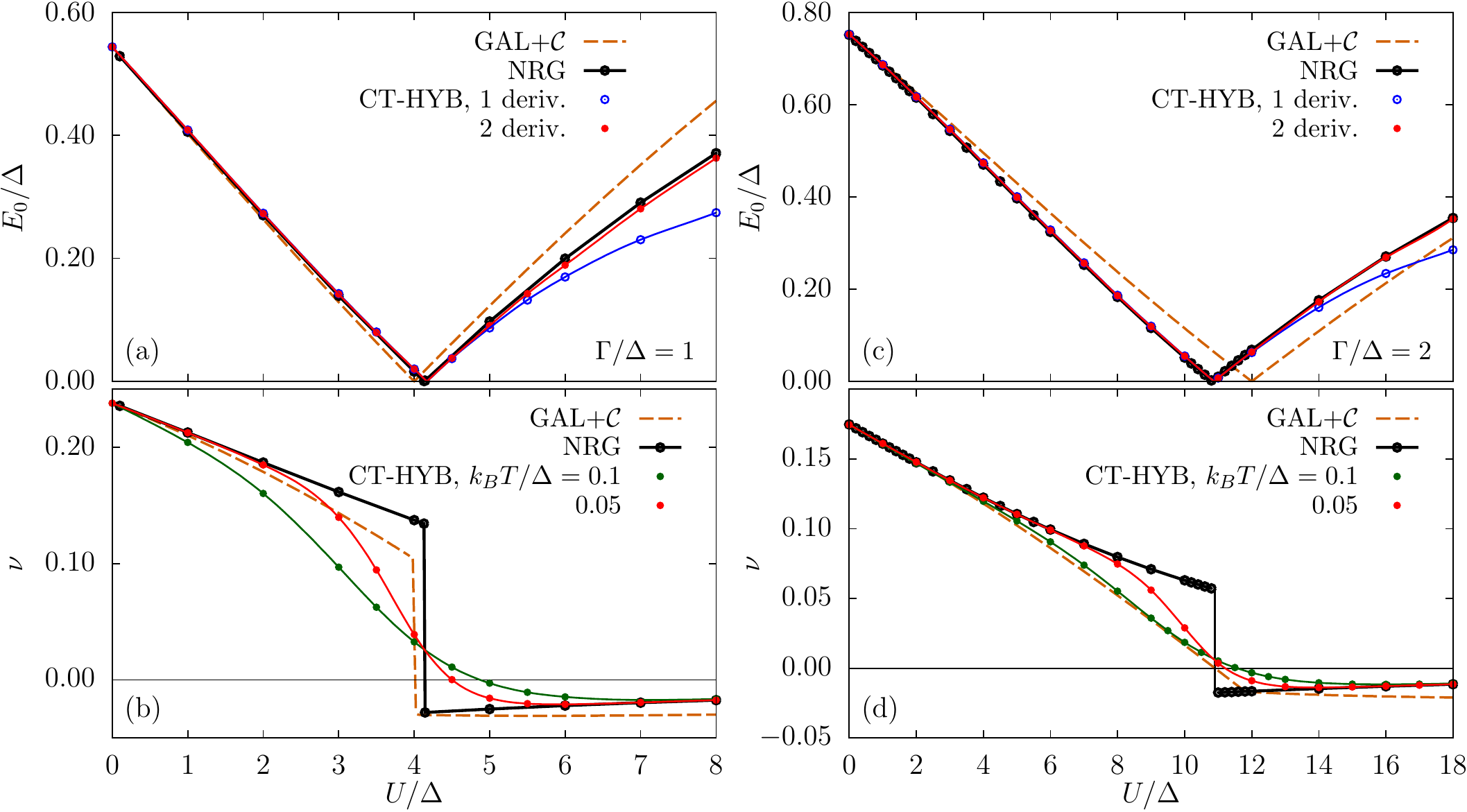}
\caption{The ABS energy $E_0/\Delta$ (top) and the induced pairing $\nu$ (bottom) as functions of the interaction strength $U$ calculated for $\varepsilon=0$ (half-filling), $\varphi=0$ and two values of the coupling strength $\Gamma=\Delta$ (panels a and b) and $\Gamma=2\Delta$ (panels c and d). Black bullets represent the NRG solution at $T=0$, orange dashed line is the GAL+$\mathcal{C}$ result. Blue and red bullets in panels (a) and (c) represent CT-HYB solution calculated at $k_BT=0.05\Delta$ using just the first derivative (blue) and first two derivatives (red) of the self-energy. Green and red bullets in panels (b) and (d) are the CT-HYB solution calculated at $k_BT=0.1\Delta$ (green) and $0.05\Delta$ (red). Lines are splines of CT-HYB data and serve only as guides for the eye. QMC error bars are smaller than the symbol size. \label{Fig:Udep}}
\end{figure*}

The Coulomb interaction strength $U$ is usually the dominant energy scale in realistic superconducting quantum dots and its value dictates much of their behavior. In particular, large values of $U$ prohibit the double occupancy of the impurity level and can drive the system into the $\pi$ phase with the doublet ground state.

In Fig.~\ref{Fig:Udep}(a) we plotted the positive ABS energy $E_0/\Delta$ together with the induced pairing $\nu$ as functions of the interaction strength $U$ at half-filling and $\varphi=0$ for two values of the tunneling rate $\Gamma=\Delta$ and $\Gamma=2\Delta$. Panels (a) and (c) show the comparison of the ABS energy calculated using NRG, CT-HYB and GAL+$\mathcal{C}$. Blue and red bullets represent CT-HYB solution at finite temperature $k_BT=0.05\Delta$ calculated using only the first derivative of the self-energy (blue) and the first two derivatives (red), respectively. The agreement with the NRG result calculated at $T=0$ is almost perfect in the $0$ phase and in the vicinity of the QPT. The effect of the second derivative of $\Sigma$ is visible only in the $\pi$ phase at higher values of $U$ where the system becomes more correlated. Even there the low-energy model provides us with a very good result already at finite temperature and by using only the first two terms in the expansion series~\eqref{Eq:LowFreqExp}, for which we need to know the self-energy only at the first two positive Matsubara frequencies $\omega_0=\pi k_BT$ and $\omega_1=3\pi k_BT$.

Panels (b) and (d) show the behavior of the induced pairing $\nu$. The green and red bullets represent the CT-HYB result calculated from the impurity density matrix at two temperatures $k_BT=0.1\Delta$ and $0.05\Delta$. These values converge to the NRG result with decreasing temperature, but more slowly than the ABS energy, still showing sizable differences at $k_BT=0.05\Delta$ in the vicinity of the transition point, while the ABS energy is already in good agreement.

In all panels of Fig.~\ref{Fig:Udep} the GAL+$\mathcal{C}$ result at $T=0$ (orange dashed lines) are shown as well. Its agreement with the NRG ABS energies is, considering the simplicity of the GAL effective model, reasonable. In general the GAL+$\mathcal{C}$ provides better predictions for lower values of the tunneling rate $\Gamma$. This is understandable taking into account the atomic nature of GAL, which becomes exact in the limit $\Gamma\rightarrow 0$. On the other hand, panels (b) and (d) show that the GAL+$\mathcal{C}$ value of the induced gap $\nu$ is not reliable at larger values of $U$ and $\Gamma$. The reason behind this is that the atomic limit gives only a trivial result on the induced pairing which is at half filling either $1/2$ in the $0$ phase or zero in the $\pi$ phase, independent of the model parameters. Therefore, its dependence on $U$ is given only indirectly by the effect of the band correction $\mathcal{C}$ on the total Green function. Nonetheless, the induced gap cannot be measured directly in experiments and, as we discuss in the next subsection, the experimentally relevant Josephson current is captured correctly. Therefore, the incorrect predictions of the induced gap do not diminish the usefulness of the GAL+$\mathcal{C}$ approximation.

\subsection{Effect of temperature \label{SSec:ResTemp}}

\begin{figure}[htb]
\includegraphics[width=\columnwidth]{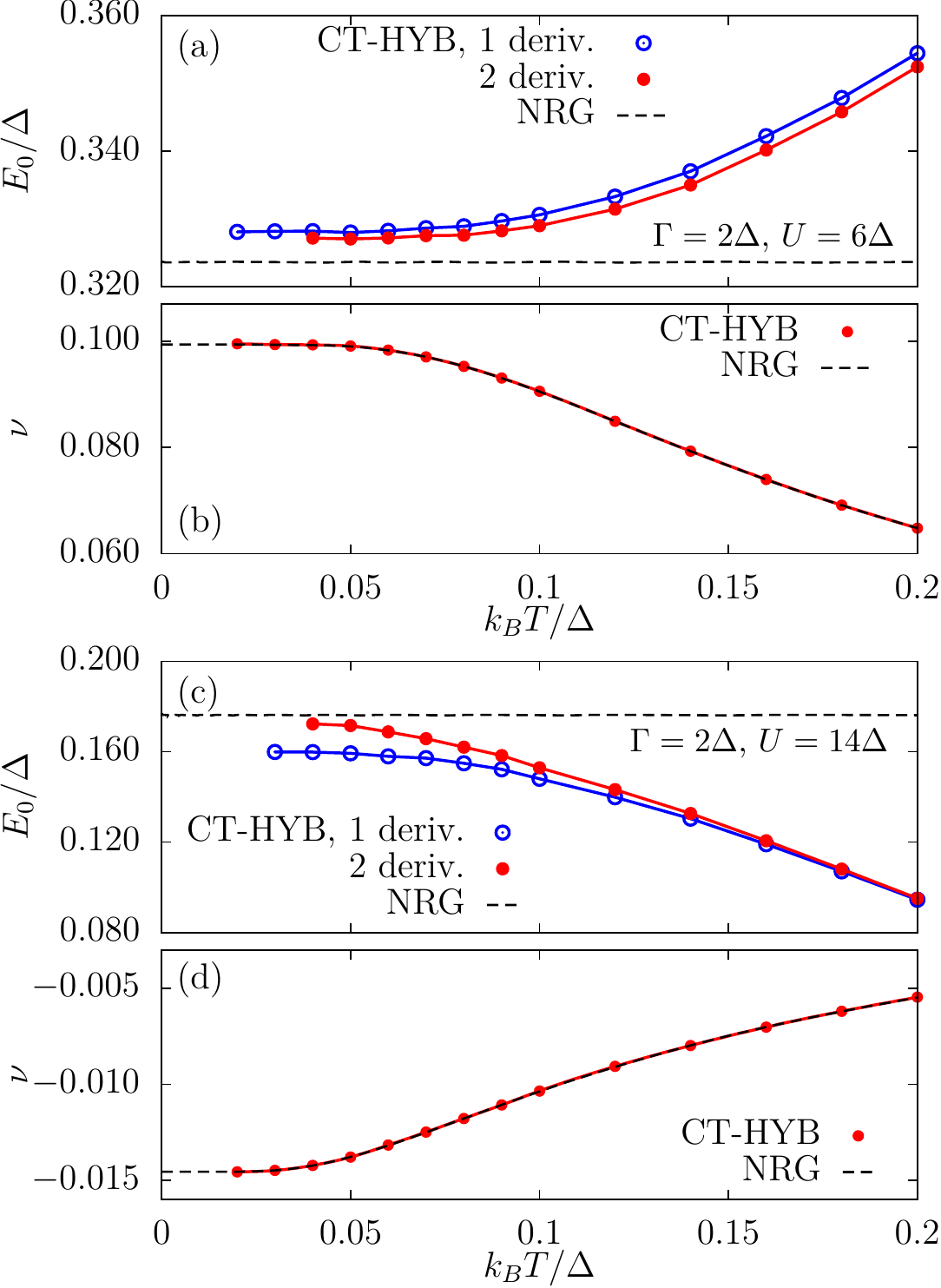}
\caption{The ABS energy $E_0/\Delta$ and the induced pairing $\nu$ as functions of the temperature calculated for $\Gamma=2\Delta$, $\varepsilon=0$ (half-filling), $\varphi=0$ and two values of the interaction strength $U=6\Delta$ (0 phase) and $U=14\Delta$ ($\pi$ phase) using CT-HYB and NRG. Blue (red) bullets in panels (a) and (c) represent the CT-HYB solution, which utilizes the first (first two) derivatives of the self-energy, black dashed line is the finite-temperature NRG solution. QMC error bars are smaller than the symbol size. The CT-HYB values of $E_0$  are not plotted for the lowest temperatures as the accuracy of the calculated derivatives of the self-energy is too low.
\label{Fig:TdepU6}}
\end{figure}

The CT-HYB is an inherently finite-temperature method with rather unfavorable scaling of the computational time with decreasing temperature as $t_c\sim T^{-2}$, which prohibits us from accessing the low-temperatures regime. Therefore, it is viable to assess how strong is the effect of the temperature on the ABS energy extracted using the presented low-energy model, i.e., how low temperatures are needed for a sufficiently precise extraction of ABS. To illustrate this, we plotted in Fig.~\ref{Fig:TdepU6} the temperature dependence of the ABS energy together with the induced pairing calculated using CT-HYB and finite-temperature NRG for $\Gamma=2\Delta$, $\varepsilon=0$, $\varphi=0$ [same as in Fig.~\ref{Fig:Udep}(c)] and two values of the interaction strength $U=6\Delta$ (0 phase) and $U=14\Delta$ ($\pi$ phase).

The ABS energy extracted from the CT-HYB in the $0$ phase is practically stable below $k_BT\approx0.1\Delta$ showing only a slight increase with increasing temperature. Above this temperature the value starts to change more rapidly. We observe an opposite trend in the $\pi$ phase. There is a slight decrease of the ABS energy below $k_BT\approx0.1\Delta$ followed by a more significant drop for larger temperatures. The observed increase in the $0$ phase and decrease in the $\pi$ phase is qualitatively consistent with previous studies of the evolution of the subgap states, e.g., the NRG results for a Kondo impurity in superconducting medium~\cite{Liu-2019} or the perturbation theory results for SCIAM~\cite{Janis-2022}. However, as already discussed, the low temperature ($k_BT\lesssim0.1\Delta$) CT-HYB results show only a very weak temperature dependence. In this respect the low-energy model is in agreement with finite-temperature NRG calculations. These predict an ABS energy practically independent of the temperature in the whole investigated range in agreement with previous NRG results of \v{Z}itko in Ref.~\cite{Zitko-2016}. As a result, the ABS energies extracted from CT-HYB data for $k_BT=0.05\Delta$ plotted in Fig.~\ref{Fig:Udep} are already in a very good agreement with the zero-temperature NRG results. Consequently, for practical purposes there is no need to perform a computationally much expensive calculation at lower temperatures.

Still, the ABS energy from CT-HYB shows at low temperatures a small offset (less than 2\%) compared to the NRG. This is partially due to the missing contributions from higher-order derivatives of the self-energy, but we cannot rule out a small systematic discrepancy between the methods. This is surprising as the induced gap, plotted in panels (b) and (d), matches the NRG value up to four decimal places, showing a remarkable agreement between the two methods over the whole temperature range.

We also note that while our low-energy model always gives sharp ABS, in reality the subgap peaks in the spectral function have non-zero width and also asymmetric shape at finite temperatures. This is due to intra-band transitions above the gap edge $\Delta$, which form a broader peak around the ABS energy as discussed in detail in Ref.~\cite{Zitko-2016}. Such intra-band transitions are beyond the realm of our atomic-like low-energy model, which always predicts sharp ABS. However, they might contribute to the observed shift of the ABS energy with increasing temperature as they have an effect on the CT-HYB self-energy. This skews the results of the low-energy model and limits its usability to lower temperatures $k_BT\lesssim0.1\Delta$.

\subsection{Current-phase relation \label{SSec:CPR}}

In contrast to the interaction strength, which is a material property, phase difference can be tuned in generalized SQUID setups by applied magnetic field~\cite{Cleuziou-2006,Delagrange-2015}. The non-zero phase difference is then the source of the equilibrium, dc Josephson current $J$ flowing between the two superconducting leads. The current-phase relation $J(\varphi)$ is an important and experimentally accessible characteristic of any superconducting junction~\cite{Golubov-2004}.

\begin{figure}[htb]
 \includegraphics[width=\columnwidth]{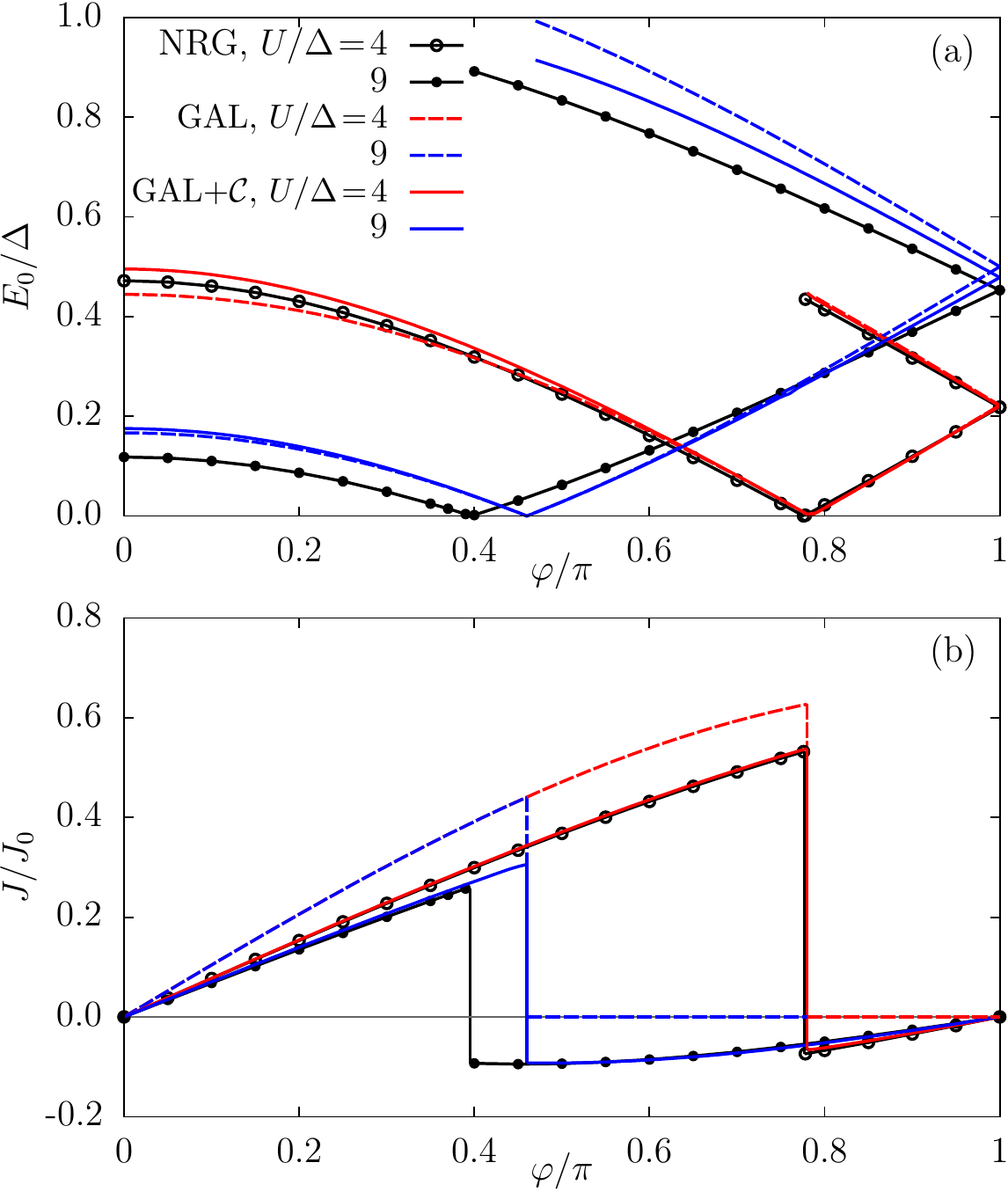}
 \caption{ABS energy $E_0/\Delta$ (a) and normalized Josephson current $J/J_0$ ($J_0=e\Delta/\hbar$) (b)  as functions of phase difference $\varphi$ for $\Gamma=2\Delta$ at half filling and two values  of the interaction strength $U=4\Delta$ and $U=9\Delta$ calculated  using NRG (black bullets), GAL without the band correction $\mathcal{C}$ (dashed lines)  and GAL with the correction (solid lines). The current calculated without  the correction is independent of the interaction strength in the $0$ phase  and is zero in the $\pi$ phase. Both these drawbacks are cured by the band correction.
 \label{Fig:PdepG2}}
\end{figure}

In Fig.~\ref{Fig:PdepG2} we plotted the positive ABS energy $E_0/\Delta$ and the dc Josephson current $J/J_0$ as functions of the phase difference $\varphi$ for $\varepsilon=0$, $\Gamma=2\Delta$ and two values of the interaction strength, $U=4\Delta$ and $U=9\Delta$. As the CT-HYB results on the Josephson current were already discussed elsewhere~\cite{Domanski-2017,Kadlecova-2019}, we plot only the result of the GAL, calculated both with and without the band correction $\mathcal{C}$, compared to the NRG result. Panel (a) shows that the band correction causes only a slight shift of the ABS energies as discussed in detail in Appendix.~\ref{App:BandCorr}, which are in both cases in a rather good agreement with the NRG.

The effect of the band correction is much more pronounced in panel (b), which shows the current-phase relation $J(\varphi)$. The dashed lines represent the bare GAL solution without band correction $\mathcal{C}$. Because the bare GAL is basically just the atomic limit with scaled parameters, we can illustrate on these results some of the more serious qualitative problems of the superconducting atomic limit when compared to the precise NRG solution. As already mentioned in Sec.~\ref{Sec:Method}, the Josephson current in this limit can be calculated as a derivative of the ground state energy w.r.t. the phase difference.
In $0$ phase it reads
$J=J_0\tilde{\Gamma}^2\sin(\varphi)/(2\sqrt{\tilde{\Gamma}_\varphi^2+\tilde{\varepsilon}^2})$
and is independent of the interaction strength $U$. Furthermore, as the ground state energy in the $\pi$ phase is independent of the phase difference, the current there is trivially zero.

The GAL+$\mathcal{C}$ results (solid lines) were calculated using formula~\eqref{Eq:JC} which correctly incorporates the effects of the bands. They show that the band correction $\mathcal{C}$ is a sufficient remedy for both qualitative problems. It reintroduces the $U$-dependence of the current in the $0$ phase and is also the source of the negative current in the $\pi$ phase. Moreover, the current is also in quantitative agreement with the NRG results, the only clear distinctions being the shift of the position of the QPT for larger $U$ as already seen in the ABS profile. The importance of such band corrections to the supercurrent were already discussed in literature for Josephson junctions with metallic (SNS) and insulating (SIS) barriers~\cite{Kuplevakhskii-1991,Chang-1994}. Our analysis proves their importance also for functionalized, S-QD-S junctions, i.e., calculating the current only from the behavior of the ABS energies can lead to incorrect results.

Moreover, besides being a fast and reliable approximation, the GAL+$\mathcal{C}$ also provides important insight into the properties of the SCIAM. Clearly, the negative current in the $\pi$ phase is solely a result of the band correction $\mathcal{C}$ as GAL without this correction gives zero current. Considering the almost perfect agreement of GAL+$\mathcal{C}$ with the NRG in this phase, we can assume that this scenario is not different in the full numerical solution. In other words, the total contribution of the ABS to the current in the $\pi$-phase is negligible as for non-zero phase difference there are always two ABS states in this regime and their contributions cancel each other out.

\subsection{Effect of local energy level \label{SSec:ResEps}}

\begin{figure*}[htb]
\includegraphics[width=2\columnwidth]{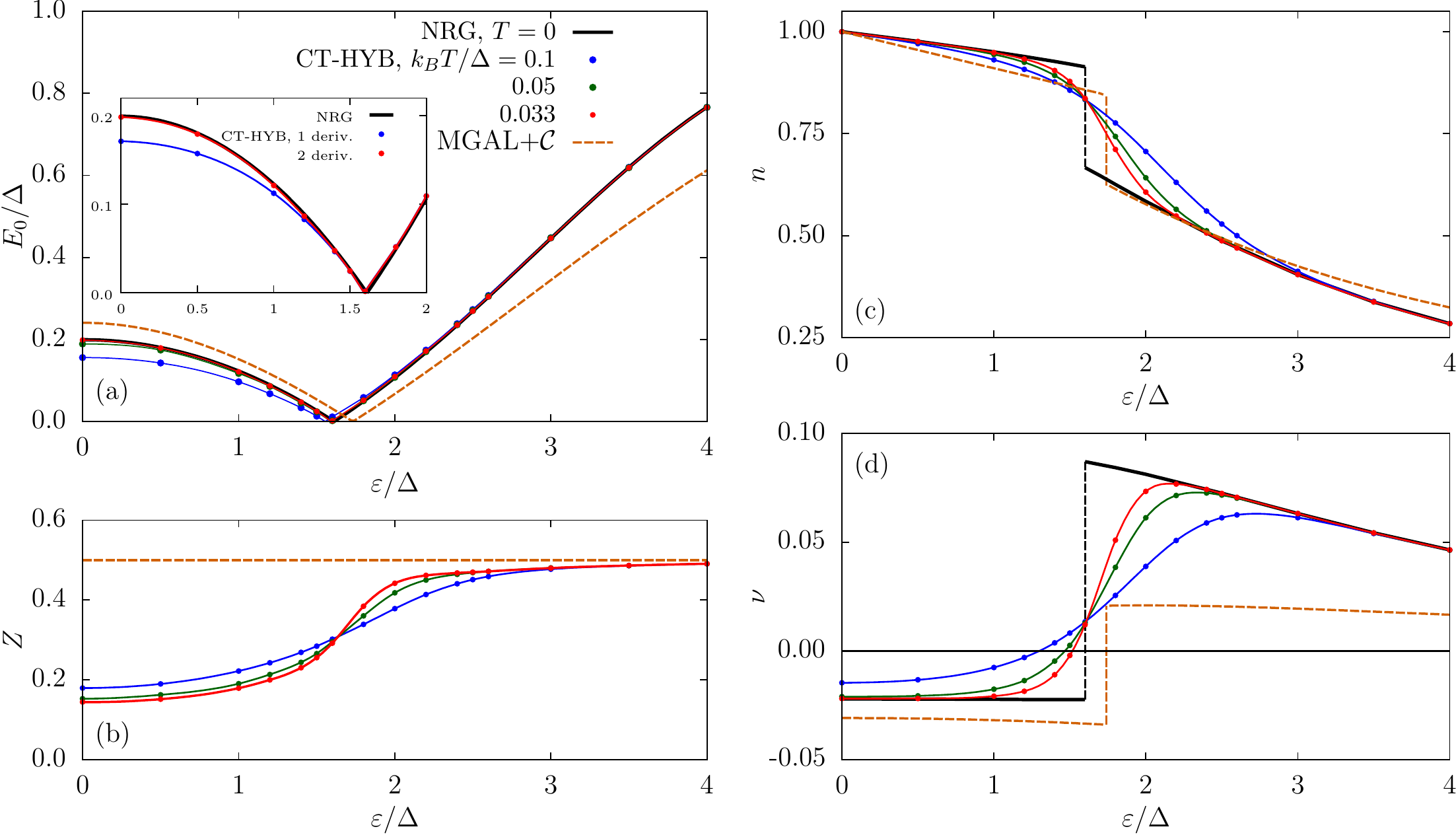}
\caption{The ABS energy $E_0/\Delta$ (a), renormalization factor $Z$ (b), total charge $n$ (c) and the induced pairing $\nu$ (d) as functions of the local energy level $\varepsilon=\varepsilon_d+U/2$ calculated using NRG at $T=0$ (solid black lines), CT-HYB at $k_BT=0.1\Delta$ (blue bullets), $0.05\Delta$ (green bullets) and $0.033\Delta$ (red bullets) and MGAL+$\mathcal{C}$ at $T=0$ (orange dashed lines). The inset shows the difference between the ABS energy calculated by CT-HYB at $k_BT=0.05\Delta$ using only the first derivative (blue) and using two derivatives (red) of the self-energy compared to NRG (black). The parameters are $U=6\Delta$, $\Gamma=\Delta$ and $\varphi=0$. Lines are splines of CT-HYB data and serve only as guides for the eye. QMC error bars are smaller than the symbol size.
\label{Fig:epsdepU6}}
\end{figure*}

So far we have discussed only the half-filled case. Now we turn our attention to the
effect of the local energy level. To address its influence is important because this
parameter can be easily tuned in experimental setups by gate voltage.
In Fig.~\ref{Fig:epsdepU6}(a) we show the dependence of the positive ABS energy $E_0$
on the local energy level $\varepsilon$ for $U=6\Delta$, $\Gamma=\Delta$ and $\varphi=0$.
Here we compare the NRG results at $T=0$ (solid black line) with the CT-HYB results at
three different temperatures $k_BT=0.1\Delta$ (blue), $0.05\Delta$ (green) and $0.033\Delta$ (red)
to further assess the convergence with the decreasing temperature. The CT-HYB calculations
utilize the first two derivatives of the self-energy. The inset shows for
$k_BT=0.033\Delta$ the comparison with the simplified method (blue circles),
which takes into account only the first derivative. The system is in $\pi$ phase
at half filling, with increasing $\varepsilon$ both NRG and CT-HYB predict the QPT at
$\varepsilon\approx1.60\Delta$. Above this value, i.e., in the $0$ phase, the
agreement between CT-HYB and NRG is almost perfect no matter the temperature.
In the $\pi$ phase there are obvious difference, however, the CT-HYB results clearly
converge to the NRG with the decreasing temperature. The relative difference between these
two methods is for $k_BT=0.05\Delta$ within 5\% and for $k_BT=0.033\Delta$ it drops below 0.5\%.
For completeness, we also provide the result of the MGAL+$\mathcal{C}$ method result at
$T=0$ (dashed orange line). It provides a reasonable quantitative estimate of the ABS
with critical point located at $\varepsilon\approx1.75\Delta$.

In panel (b) we show the value of the renormalization factor
$Z=[1+\Gamma/\Delta-\Sigma^\prime(0)]^{-1}$ from CT-HYB together with
the value of $q$ from MGAL to illustrate the effect of the first
derivative of the self-energy. The renormalization is strongest at
half-filling and decreases rapidly in the vicinity of the transition point.
The derivative $\Sigma_n^\prime(0)$ is very small in the $0$ phase, similarly
to the solution in the atomic limit. For $\varepsilon\rightarrow\infty$
then $Z$ approaches the MGAL value $q=(1+\Gamma/\Delta)^{-1}=1/2$.

Panels (c) and (d) illustrate the effect of the position of the energy level on the total
charge $n$ and the induced pairing $\nu$. The CT-HYB results approach the NRG result with
decreasing temperature as expected. The MGAL also provides a reasonable estimate of the total
charge $n$. On the other hand, the value of the induced pairing is again off as it is much
lower than the exact result, for reasons already explained in the previous section.

We note that the lines calculated using CT-HYB at different temperatures cross at the same point, which coincides with the position of the $T=0$ QPT. This behavior can be understood by a mapping of SCIAM to a two-level model which proves that at low enough temperatures (ca. $k_BT<0.1\Delta$) all physical observables become temperature-independent at the QPT, as explained in detail in Refs.~\cite{Kadlecova-2019,Pokorny-2021}. This feature can be utilized to locate the transition point from finite-temperature QMC or experimental data.

\subsection{The fate of the second pair of ABS\label{SSec:Res2ndABS}}
The main disadvantage of the low-energy model~\eqref{Eq:GreenLowEn} is its inability to provide results on the second pair of ABS, which may be present in the $\pi$ phase. Even the extended model~\eqref{Eq:ExtLowEnergy}, which takes two energy derivatives into consideration, always predicts a second state well above the gap edge $\Delta$. Here we show that this limitation is not severe as the region of the parameter space where the spectral function contains two pairs of ABS and simultaneously the second ABS is recognizable in practical realizations is small.

\begin{figure}[htb]
\includegraphics[width=\columnwidth]{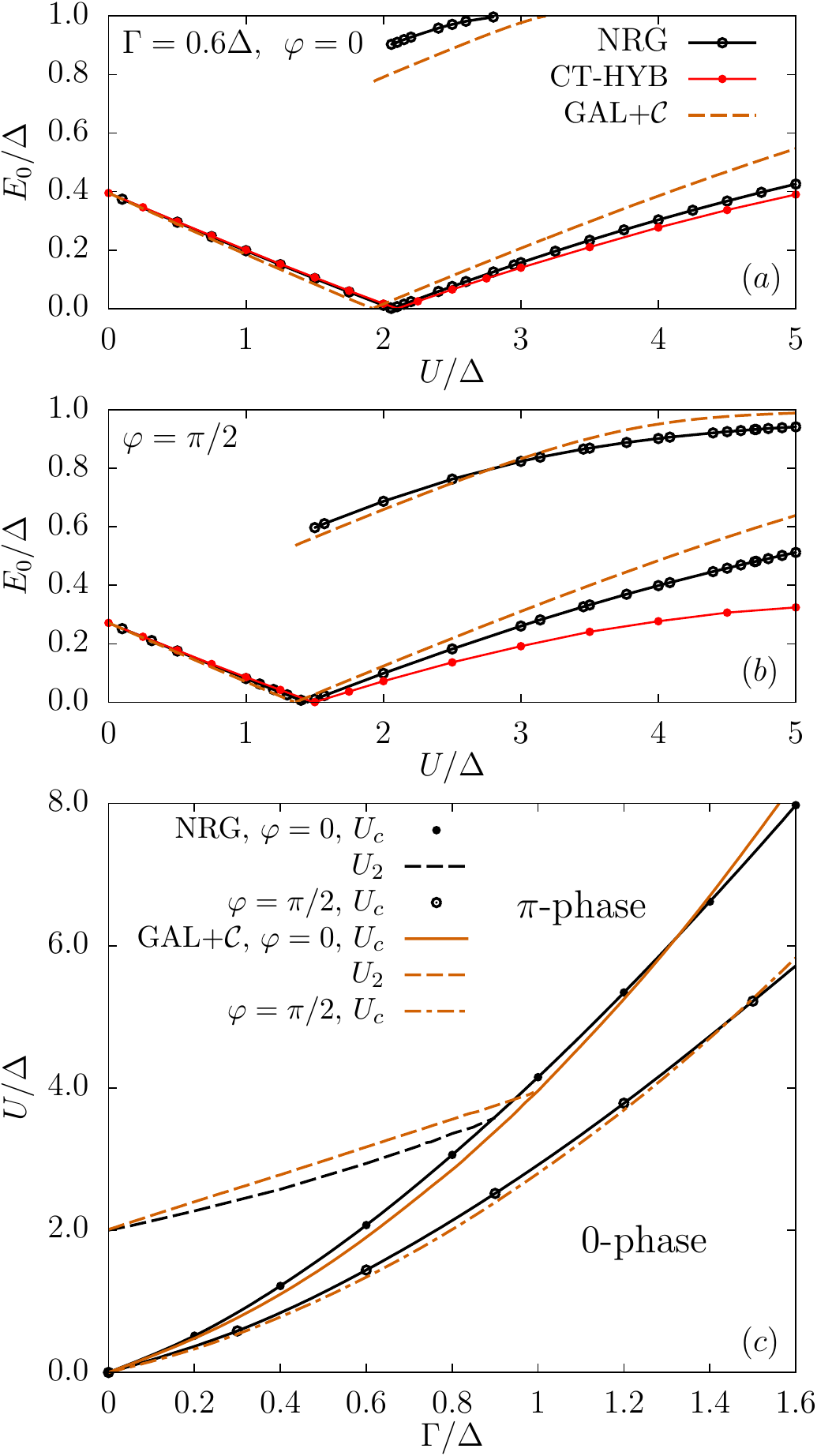}
\caption{(a) ABS energies as functions of the interaction strength for $\Gamma=0.6\Delta$ at half-filling and $\varphi=0$ calculated using NRG at $T=0$ (black bullets), CT-HYB at $k_BT=0.05\Delta$ (red bullets) and GAL at $T=0$ (dashed lines). (b) Same plot for $\varphi=\pi/2$. (c) Phase diagram of SCIAM at zero temperature and half-filling in the $\Gamma-U$ plane for $\varphi=0$ and $\varphi=\pi/2$. Black solid lines represent the phase boundary calculated using NRG, black dashed line is the interaction strength $U_2$ at which the second ABS vanishes from the spectral function for $\varphi=0$. Orange lines represent the GAL result.\label{Fig:phase_GU}}
\end{figure}

In Figs.~\ref{Fig:phase_GU}(a) and~\ref{Fig:phase_GU}(b) we show the positive ABS energy $E_0$ as function of the interaction strength $U$ calculated using NRG and GAL+$\mathcal{C}$ at $T=0$ and CT-HYB at $k_BT=0.05\Delta$ at half-filling, $\Gamma=0.6\Delta$ and two values of phase difference $\varphi=0$ and $\varphi=\pi/2$. The main difference between the case of zero and non-zero phase difference is the fate of the outer ABS with increasing interaction strength. For zero phase difference, both NRG and GAL+$\mathcal{C}$ results project that the second ABS will vanish into the continuum above some interaction strength. In particular, for $\varphi=0$ the NRG predicts the second ABS to emerge at the phase transition point ($U_c\approx2.05$) and to enter the continuum at $U_2\approx2.85$. The non-zero phase difference promotes the $\pi$ phase and the phase boundary is shifted to lower values, for $\varphi=\pi/2$ we get $U_c\approx1.40$. More importantly, the second ABS does not vanish from the gap but continuously approaches the gap edge, until it becomes indistinguishable from the continuum~\footnote{
The difference in the behavior of the ABS for zero and non-zero $\varphi$ can be proven
by calculating numerically the derivative $\partial E_0/\partial U$, which changes
abruptly at $U=U_2$ for $\varphi=0$ but continuously approaches zero for $\varphi=\pi/2$.
}.

GAL$+\mathcal{C}$ provides a good quantitative estimate of the development of both states as the chosen value of the tunneling rate is rather small. CT-HYB, on the other hand, fails to recognize the second ABS. The agreement between CT-HYB and NRG also worsens as we enter deeper into the $\pi$ phase where the effects of the higher order derivatives of the self-energy become significant.

Fig.~\ref{Fig:phase_GU}(c) shows the phase diagram of SCIAM at zero temperature and half-filling in the $\Gamma-U$ plane for $\varphi=0$ and $\varphi=\pi/2$. Solid lines represent the phase boundary $U_c$ at $\varphi=0$ and the dashed lines mark the value $U_2$ where the second ABS vanishes for $\varphi=0$. We see that for $\varphi=0$ the existence of the second ABS is bound to low values of the interaction strength, far below the experimental range of $U/\Delta\sim10$. Therefore, the inability of the low-energy model to recognize its presence is not relevant for real-world systems. In case of non-zero phase difference the second ABS does not vanish but quickly becomes indistinguishable from the band.

\section{Conclusions \label{Sec:Conclusions}}

Understanding the behavior of ABS in complex nanostructures involving correlated quantum dots connected to superconducting electrodes is a crucial step towards their future applications. Here we presented two methods, which can be used for this purpose. While we tested them on the simplest setup of a single quantum dot with two superconducting leads, both methods can be straightforwardly generalized to multi-level systems. The CT-HYB calculation is computationally demanding, however it can provide unbiased results on the occupation numbers and the Josephson current up to large interaction strengths. The applicability of the presented mapping to the low-energy model, which is used to extract the ABS energies, is for now limited to the intermediate interaction strengths. The main reason is the truncation of frequency expansion of the self-energy at the second order while the higher orders become more relevant with increasing interaction strength. Here, the technical issue is that extracting higher-order derivatives from stochastic QMC data can be unreliable as evaluating the self-energy from the Dyson equation requires calculating the difference between the inverses of two Green functions, a procedure highly susceptible to the numerical noise. While it is possible to obtain better results by measuring higher-order correlation functions, which are related to the self-energy by the equation of motion~\cite{Hafermann-2012}, this option is not implemented in the TRIQS solver. Another possibility would be to utilize alternative representations of the Green function, e.g., the expansion in the basis of Legendre polynomials~\cite{Boehnke-2011} which act as an effective noise filter. Yet, we would like to stress that the truncation to the second order is already able to give sufficiently precise predictions for the position of the ABS for experimentally relevant parameters, especially, when compared to the finite accuracy of a typical experiment.

On the other hand the GAL method with here introduced band corrections is computationally inexpensive. Its main limitation is that it is based on the mapping to the superconducting atomic limit, which largely ignores the presence of the leads. Therefore, GAL is bound to lower values of the dot-lead coupling. However, in this regime it gives surprisingly good estimates of any relevant measurable quantities and provides important insight into the properties of the SCIAM. Therefore, even the 'hand-waving' derivation of GAL as presented in our paper gives a very simple and fast solver for the SCIAM, which can deliver reliable results to the experimentally relevant range of parameters within seconds on a standard PC. As such, it can replace other simple methods like the second-order perturbation theory, which cannot be employed for degenerate ground states~\cite{Zonda-2015}.

Yet, strictly speaking, the method still lacks a proper formal derivation. We believe that it could be performed by the analysis of the analytical structure of the two-particle functions, similarly to the case of the Fermi liquid theory~\cite{Nozieres-1962,Luttinger-1962}. Such calculation could shed light on the peculiar MGAL scaling of the local energy level as well as provide solid argumentation on the scaling of the interaction strength. Its importance can be stressed by the fact that we have already utilized GAL to treat multi-level systems~\cite{Zonda-2023} and that GAL is considered as a fast approximate solver for the superconducting dynamical mean-field theory.

We also discussed some non-trivial insights into the physics of SCIAM, which were obtained using the two methods. The low-energy model shows a temperature dependence where the ABS energy is increasing in the $0$ phase and decreasing in the $\pi$ phase with increasing temperature. This is consistent with results of some other theoretical works~\cite{Liu-2019,Janis-2022} but it is in disagreement with the NRG results~\cite{Zitko-2016}. We also proved that the negative Josephson current in the $\pi$ phase is a result of the presence of the bands as the contributions from the two pairs of ABS cancel each other out in this phase. The behavior of the second pair of ABS in the $\pi$ phase was also studied, showing that their existence for zero phase difference is bound to low values of the tunneling rate and the interaction strength, while for non-zero phase difference they become almost indistinguishable from the band at realistic values of the interaction strength.

To summarize, we have presented two low-energy models that have proven efficient for calculating ABS energies of superconducting impurity systems. GAL provides a fast and reasonably accurate approximative solution of the SCIAM and is suited for extensive parameter scans, e.g., for an initial analysis of experimental data or as a starting point of more elaborated calculations. On the other hand, the effective mapping to a low-energy model that allows the extraction of the Andreev bound state energies from unbiased imaginary-time quantum Monte Carlo simulations can be employed to obtain precise results for realistic setups. Together, these techniques represent an efficient toolbox for modeling realistic nanoscopic superconducting devices.

\begin{acknowledgments}
We acknowledge fruitful discussions with T. Novotn\'y and P. Zalom.
This research was supported by the project e-INFRA CZ (ID:90140) of the Czech Ministry of Education, Youth and Sports,
by Grant No. 22-22419S (M.\v{Z}) of the Czech Science Foundation and
the COST action SUPERQUMAP (CA21144).
\end{acknowledgments}

\appendix

\section{Scaling of the interaction strength in the GAL scheme\label{App:GALcalc}}
Here we provide the derivation of the scaling of the interaction strength $U$ and the value of the shift of the chemical potential in the GAL scheme, Eq.~\eqref{Eq:GAL_U}. For sake of simplicity we neglect the band correction $\mathcal{C}$ as it does not change the result. The relation between the GAL self-energy and the atomic self-energy from the auxiliary atomic problem~\eqref{Eq:HamGALFull} reads $\Sigma(\omega)=q^{-1}\tilde{\Sigma}_\infty(\omega)$. In the $0$ phase the self-energy in the atomic limit is static and its normal and anomalous components follow Eq.~\eqref{Eq:SEatS}, from which we obtain a pair of equations reading
\begin{equation}
\label{Eq:n_vs_tilden}
Un+\mu_s=q^{-1}\tilde{U}(\tilde{n}-1/2),\quad
U\nu=q^{-1}\tilde{U}\tilde{\nu}.
\end{equation}
The electron density and the induced pairing can be calculated from the respective Green functions as
\begin{equation}
\begin{aligned}
n&=-\frac{2}{\pi}\int_{-\infty}^{\infty}d\omega f(\omega)\Imm G_n(\omega+i0), \\
\nu&=-\frac{1}{\pi}\int_{-\infty}^{\infty}d\omega f(\omega)\Imm G_a(\omega+i0),
\end{aligned}
\end{equation}
where $f(\omega)=[e^{\omega/k_BT}+1]^{-1}$ is the Fermi-Dirac distribution. Similar relations bind the quantities $\tilde{n}$ and $\tilde{\nu}$ to the normal and anomalous elements of the auxiliary Green function $\tilde{G}(\omega)$. The relation between the two Green functions is given by Eq.~\eqref{Eq:GFgalFull}, $G(\omega)=q\tilde{G}(\omega)$, from which we obtain the relations between the occupation numbers, $\nu=q\tilde{\nu}$ and $n=q\tilde{n}$. This also illustrates the problem of the missing spectral weight in GAL without the band correction: The auxiliary atomic problem at half-filling ($\tilde{n}=1$) corresponds to GAL at filling $n=q<1$. Inserting these relations in Eq.~\eqref{Eq:n_vs_tilden} we obtain $\tilde{U}=q^2U$ and $\mu_s=-qU/2$.

While this scaling was derived only for the $0$ phase, we use the same scaling also in the $\pi$ phase, where such simple argumentation is not possible due to the frequency dependence of the self-energy. The reason for this is that using different scaling in the two phases would result in disagreement about the position of the QPT while approaching it from each phase.

\section{Effects of the band correction\label{App:BandCorr}}

\begin{figure}[]
\includegraphics[width=\columnwidth]{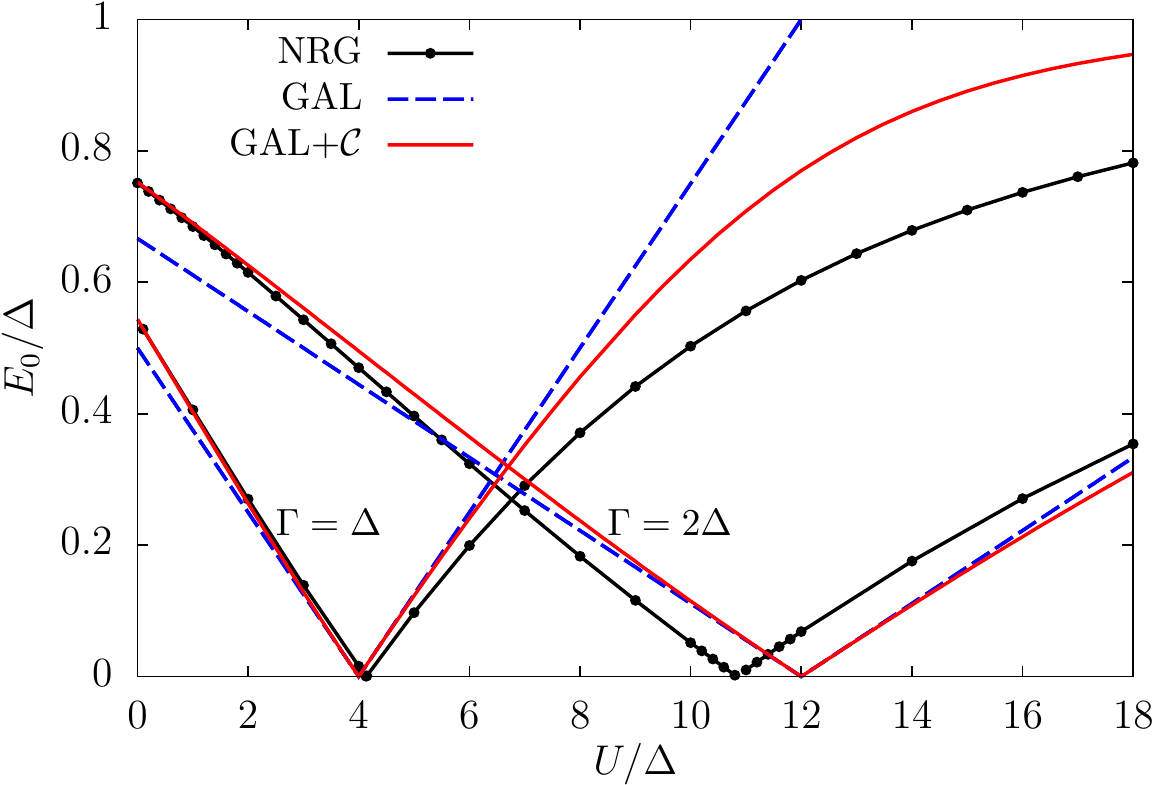}
\caption{ABS energy $E_0/\Delta$ as a function of the interaction strength $U$ for the same parameters as in Fig.~\ref{Fig:Udep}: $\varepsilon=0$, $\varphi=0$ and two values of the tunneling rate $\Gamma=\Delta$ and $\Gamma=2\Delta$, calculated using GAL including the band correction $\mathcal{C}$ (solid red), without the correction (dashed blue) and NRG (black). The effect of the band correction is stronger at larger values of $E_0$ and vanishes exactly at the transition point ($E_0=0$).
\label{Fig:UdepCorr}}
\end{figure}

Let us discuss further the importance of the band correction $\mathcal{C}$, Eq.~\eqref{Eq:Corr}. As the correction is the same in both presented methods, we resort to GAL results only. In Fig.~\ref{Fig:UdepCorr} we plotted the positive ABS energy as a function of the interaction strength $U$ for the same parameters as in Fig.~\ref{Fig:Udep}. Solid red (dashed blue) lines represent the GAL solution with
(without) the correction, compared to the NRG result (black bullets). As the correction vanishes for $\omega\rightarrow 0$, it cannot change the position of the QPT, which is given solely by the zero of the right-hand side of Eq.~\eqref{Eq:ABSbareCTHYB}. Its effects become stronger with increasing values of $E_0/\Delta$ where it prevents the ABS from entering the continuum, keeping its energy below $\Delta$. It also guarantees that GAL becomes exact in the non-interacting ($U=0$) limit. Furthermore, it corrects the high-frequency asymptotics of the GAL Green function, as already mentioned in Sec.~\ref{SSec:GAL} and it has an important
contribution to the Josephson current as discussed in Sec.~\ref{SSec:ResTemp}.

%

\end{document}